\documentclass[pra,twocolumn,epsfig,rotate,superscriptaddress,showpacs]{revtex4}
\usepackage[dvips]{graphics}
\usepackage{graphicx}
\usepackage{amsfonts}
\usepackage{amssymb}
\usepackage{amsmath}
\usepackage{subfigure}

\usepackage{prettyref}
\usepackage{float}
\usepackage{amsmath}
\usepackage{amssymb}
\usepackage{esint}

\begin{document}
\title
{Quantum adiabatic evolution with energy-degeneracy levels}

\author{Qi Zhang}
\affiliation{Wilczek Quantum Center and College of Science, Zhejiang University of Technology,
Hangzhou 310014, People's Republic of China}

\date{\today}
\begin{abstract}
A classical-kind phase-space formalism is developed to address the tiny intrinsic dynamical deviation from what is predicted by Wilczek-Zee theorem during quantum adiabatic evolution on degeneracy levels. In this formalism, the Hilbert space and the aggregate of degenerate eigenstates become the classical-kind phase-space and a high-dimensional subspace in the phase-space, respectively. Compared with the previous same study by a different method, the current result is qualitatively different in that the first-order deviation derived here is always perpendicular to the degeneracy subspace. A tripod scheme Hamiltonian with two degenerate dark states is employed to illustrate the adiabatic deviation with degeneracy levels.
\end{abstract}

\maketitle

\section{Introduction}

Quantum adiabatic evolution is always of fundamental interests to physicists since the discovery of quantum adiabatic theorem~\cite{adiabatic}. It predicts the general and fundamental behaviors of a quantum system under a slow external driving. Due to the approximated nature of adiabatic theorem, intrinsic deviation from what is predicted by adiabatic theorem inevitably arises~\cite{deviation}. Although most of attention has been paid to the adiabatic deviations and conditions in the case of non-degeneracy energy spectrum, which seems theoretically more fundamental and simple~\cite{controversy}, the study on adiabatic evolution for degeneracy spectrum may be more important in a practical sense, which is more related to holonomic quantum computation and detection of fractional statistics. The study of degenerate adiabatic deviation may be closely associated with assessing the feasibility of topological gates using the concept of Majorana non-Abelian braiding~\cite{braiding}.

Because the analytical formulae about the adiabatic deviation in the case of non-degeneracy energy spectrum are already very complicated, few people has ever touched upon that for the formidable degeneracy spectrum case~\cite{degenerate1,degenerate2,degenerate3}. The difficulty mostly comes from the abstract nature of Hilbert space and the quantum-kind of formulae. In this paper, we focus on the adiabatic deviations during which the state under study is always degenerate with other orthogonal states by projecting the Hilbert space onto a phase-space of classical form and mapping the eigenstates onto fixed points in the phase-space thus defined, which simplifies greatly the analytical expressions and enables the visual comprehension of the deviations.

In the current theory the aggregate of eigenstates in the degeneracy subspace forms a patch in phase-space rather than isolated point in the non-degenerate case. The patch can be high dimensional according to the degree of degeneracy. With the overall phase of wavefunction omitted, each point on the degeneracy subspace is a fixed point of classical form of Hamiltonian. In the first-order theory with respect to slow adiabatic speed, which is of overwhelming importance, the difference between the real state and what is predicted by the Wilczek-Zee theory~\cite{WZ} is always perpendicular to the degeneracy patch (see Fig.~1), with the distance between the average of the oscillating deviation and the Wilczek-Zee point being proportional to the adiabatic speed.

In higher order formulation, the deviation may have components in the degeneracy subspace if, more intuitively and physically, the degeneracy subspace in the whole phase-space changes its normal direction during the adiabatic manipulation. We use the example of the tripod scheme, where three laser beams are interacting with a free Rubidium atom, to verify our theory. Theoretically, the tripod scheme has been introduced to implement the non-Abelian vector potential and spin-order coupling on neutral atoms~\cite{nonAbelian}.

Technically we take advantage of the classical Hamiltonian formulation
of the Schr\"odinger equation~\cite{Weinberg,HeslotPRD1985,Liu2003PRL}.
Note that this classical formulation is purely mathematical and is {\it not}
the traditional semiclassical limit $\hbar\rightarrow0$.

\begin{figure}[t]
\resizebox *{15cm}{15cm}{\includegraphics*{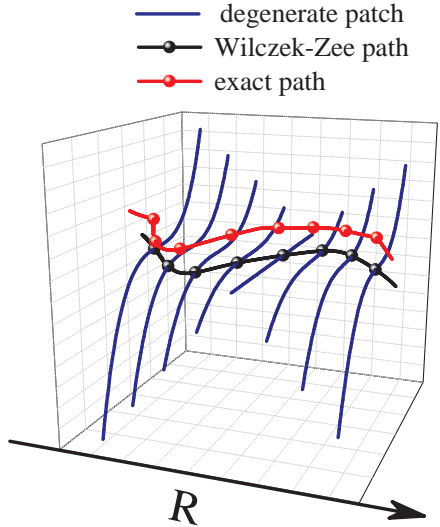}}
\vspace{-7cm} \caption{(color online) Illustration of the exact state from first-order calculation and Wilczek-Zee state on the degeneracy subspace. For convenience, the parameter-dependent degeneracy patch is demonstrated by one dimensional curve, although it is in fact a $2(m-1)$ dimensional regime. The red curve out of the patch and the black curve on the patch are the trajectory of the exact state and that of the Wilczek-Zee state during the degenerate adiabatic evolution.
According to the first-order perturbation theory, the projection of the exact state on the degeneracy subspace is just the Wilczek-Zee state, which means that the non-adiabatic deviation has zero first-order component in the degeneracy subspace. }\label{fig1}
\end{figure}

\section{General quantum dynamics in the form of classical dynamics}

We consider a quantum system described by the Hamiltonian $\hat{H}(R)$,
where $R=R(t)$ represents time-dependent parameters in an adiabatic protocol. Different from the ordinary systems, here $\hat{H}(R)$ has a discrete but degeneracy spectrum during the entire control protocol. In the case that the system is initially prepared on the degeneracy levels, the adiabatic evolution (geometric phase) can be described by the Wilczek-Zee phase~\cite{WZ}. However, it has been proven that, so long as the protocol is not executed in the mathematical limit $\dot{R} \rightarrow 0$~\cite{degenerate1,degenerate2,degenerate3}, the deviation from what is predicted by Wilczek-Zee theory should be expected.

For simplicity and concrete discussion, we assume $\hat{H}(R)$ lives in a finite $n$-dimensional Hilbert space with two eigenstates being degenerate (the generalization to higher degree of degeneracy is straightforward). Suppose $|D_1\rangle$ and $|D_2\rangle$ are the two degenerate states of $\hat{H}(R)$, any state of the form $c_1|D_1\rangle+c_2|D_2\rangle$ ($|c_1|^2+|c_2|^2=1$) would be an eigenstate. In the energy representation, a state can be expressed as $c_1|D_1\rangle+c_2|D_2\rangle+\sum_{i=3}^{n}c_i|S_i\rangle$, with $|S_i\rangle(i=3,\ldots n)$ being non-degeneracy levels.

Due to the abstractness of C-numbers in Hilbert space, we alternatively employ real quantities and the corresponding classical form of phase-space to address the quantum state in Hilbert space. In particular, in energy representation, we can define the classical phase-space point
\begin{equation} \label{Idefinition}
p_i'=\arg(c_{i+1})-\arg(c_1), \quad q_i'=|c_{i+1}|^2,
\end{equation}
with $i=1,2,\ldots,n-1$, to describe the wavefunction and the phase-space without any approximation. With the phase-space thus defined, the degeneracy region is the patch spanned by $(p_1',q_1')$ in the phase-space.

Next, let's turn to the evolution of wavefunction, which definitely satisfies the Schr\"odinger equation defined by $\hat{H}(R)$. In order to gain an insight into the dynamics in the perspective of classical form phase-space, we should express the Schr\"odinger equation via the variables $(p_i',q_i')$ instead of the initial form of wavefunction. Generally, as the unitary matrix to diagonalize $\hat{H}(R)$ is $R$ (time) dependent, it is more convenient to take advantage of a general but fixed representation, which may be more physical relevant, rather than the energy representation. In the fixed representation, a wavefunction can be expressed as
\begin{equation}
|\psi\rangle=\sum_{i=1}^{n}a_i|B_i\rangle,
\end{equation}
with $|B_i\rangle$ being the orthogonal bases in Hilbert space. The classical phase-space and phase-space point is defined by \begin{equation} \label{definition}
p_i=\arg(a_{i+1})-\arg(a_1), \quad q_i=|a_{i+1}|^2,
\end{equation}
with $i=1,2,\ldots,n-1$. $(p_i',q_i')$ and $(p_i,q_i)$ are related via a $R$-dependent canonical transformation. By construction, the Schr\"odinger equation then yields the following Hamilton's equations of motion without any approximation~\cite{Weinberg,HeslotPRD1985,Liu2003PRL},
\begin{equation} \label{classical}
\quad \frac{d p_i}{dt}=-\frac{\partial H(R)}{\partial q_i}\,,
~~~\frac{d q_i}{dt}=\frac{\partial H(R)}{\partial p_i},
\end{equation}
where the classical Hamiltonian $H(R)$ is obtained from the quantum Hamiltonian  $\hat{H}(R)$ via
\begin{equation} \label{expectation}
H(R)=\langle\psi|\hat{H}(R)|\psi\rangle\,.
\end{equation}
With the overall phase removed, the phase space in this classical formalism
is just the projective Hilbert space, which provides a clear perspective of the wavefunction and its adiabatic evolution.

One final technical comment is in order. The mapping from the wavefunction components $a_i$ to phase space variables $(p_i, q_i)$ [see  Eq.~(\ref{definition})] becomes ambiguous when any one of
the wavefunction component $a_i$ becomes zero. Fortunately, this ambiguity can be easily overcome
by adopting  a different representation to re-express the wavefunction.  For example, $a_1$ in Eq.~(\ref{definition})  is used to remove the overall wavefunction phase.  If $a_1=0$,  one can always
select another nonzero $a_i$ to carry out a similar mapping.

\section{Dynamics for the degenerate adiabatic deviation: first-order theory}

During the adiabatic evolution, the initial wavefunction on the degeneracy patch should always be on the instantaneous degeneracy patch and follow the Wilczek-Zee theory. However, as in the case of non-degenerate adiabatic following~\cite{deviation}, the deviation from what is predicted by Wilczek-Zee theory has been proven to arise~\cite{degenerate1,degenerate3}. In the language of phase-space point, the deviation should be expressed,
\begin{equation} \label{expansion}
p_i(t)=\bar{p}_i[R(t)]+\delta p_i,\ \  q_i(t)=\bar{q}_i[R(t)]+\delta q_i,
\end{equation}
with $(\delta p_i, \delta q_i)$ being time-dependent deviations from the ideal adiabatic trajectory $[\bar{p}_i(R),\bar{q}_i(R)]$ predicted by Wilczek-Zee theory. In what follows we develop the dynamics for $(\delta p_i, \delta q_i)$ to the first order of $\dot{R}$.

Employing the dynamics~(\ref{classical}) and the expression~(\ref{expansion}), the dynamics for $(\delta p_i, \delta q_i)$ can be written,
\begin{equation} \label{fd}
\left(\begin{array}{c}\frac{\partial \bar{p}(R)}{\partial R} \dot{R} + \frac{d \delta p}{dt}\\ \\
\frac{\partial \bar{q}(R)}{\partial R} \dot{R} + \frac{d \delta q}{dt}\end{array}\right)=\Gamma
 \left(\begin{array}{c} \delta p\\ \\
 \delta q\end{array}\right),
\end{equation}
where
\begin{equation} \label{Gamma}
\Gamma=\left(\begin{array}{cc}-\frac{\partial^2H_0}{\partial q\partial
p}&-\frac{\partial^2 H_0}{\partial q\partial q}
\\ \\ \frac{\partial^2H_0}{\partial p
\partial p}&\frac{\partial^2H_0}{\partial
p \partial q}
\end{array}\right)_{p=\bar{p},q=\bar{q}}
\end{equation}
is an $R$-dependent $2(n-1)\times2(n-1)$ matrix obtained from the second-order derivatives of $H(R)$. The expression $p$ ($q$) without subscript stands for the matrix stack of the whole set of $p_i$ ($q_i$), with $i=1,2,\ldots,n-1$, e.g.,
\begin{eqnarray} \nonumber
\left(\begin{array}{cc} \delta p & \delta q\end{array}\right)^T&\equiv&\left(\begin{array}{cccccc} \delta p_1 & \ldots  & \delta p_{n-1} & \delta q_1 &  \ldots &  \delta q_{n-1}\end{array}\right)^T \\
\left(\begin{array}{cc}  \bar{p} &  \bar{q}\end{array}\right)^T&\equiv&\left(\begin{array}{cccccc}  \bar{p}_1 & \ldots  & \bar{p}_{n-1} &  \bar{q}_1 &  \ldots &  \bar{q}_{n-1}\end{array}\right)^T
\end{eqnarray}
Because $\hat{H}(R)$ is degenerate, which gives rise to the null dynamics $\dot{p}_i=\dot{q}_i=0$ as long as the deviation $(\delta p_i, \delta q_i)$ is on the degeneracy patch, the matrix $\Gamma$ must be of linear dependence and the determinant $|\Gamma|=0$.

In order to associate the dynamics with the Wilczek-Zee phase, we here carry out a canonical transformation from $(p_i,q_i)$ to $(P_i,Q_i)$, i.e.,
\begin{equation} \label{conventionn}
\Lambda\equiv\left(\begin{array}{ccccccc} P_1 & Q_1 & P_2 & Q_2 & \ldots & P_{n-1} & Q_{n-1} \end{array}\right)^T,
\end{equation}
with
\begin{equation}
P_1=p_1', \quad\quad Q_1=q_1',
\end{equation}
and
\begin{equation}
\Lambda=U \left(\begin{array}{c} p\\ \\
  q\end{array}\right), \quad \bar{\Lambda}=U \left(\begin{array}{c} \bar{p}\\ \\
  \bar{q}\end{array}\right), \quad\delta\Lambda=U \left(\begin{array}{c} \delta p\\ \\
 \delta q\end{array}\right),
\end{equation}
where $U$ is a $2(n-1)\times2(n-1)$ matrix diagonalizing $\Gamma$,
\begin{equation} \label{firstdia}
\Gamma_{\text{dia}}=U\Gamma U^{-1}=\left(\begin{array}{cccccc}
0 & 0 & 0  & 0 & \ldots & 0 \\
0 & 0 & 0  & 0 & \ldots & 0 \\
0 & 0 &d_1 & 0 & \ldots & 0\\
0 & 0 & 0  &d_2& \ldots & 0 \\
\vdots & \vdots & \vdots  &\vdots& \ddots & \vdots \\
0 & 0 & 0  & 0 &  \ldots    &d_{2(n-2)}     \end{array}\right);
\end{equation}
$\delta\Lambda$ ($\bar{\Lambda}$) is a column vector
with each component being the linear combination of $\delta p_i,\delta q_i$ ($\bar{p}_i,\bar{q}_i$) ($i=1,\ldots,n-1$). The first two columns (lines) of $\Gamma_{\text{dia}}$ with null diagonal elements stand for the direction along $p_1'=P_1$ and $q_1'=Q_1$ defined in Eq.~(\ref{Idefinition}), i.e., on the degeneracy patch. Even though the nonzero diagonal elements $d_i$ ($i=1,\ldots,2(n-2)$) may generally be complex numbers, the dynamics induced by $\Gamma_{\text{dia}}$ and the corresponding vector $\Lambda$ is totally equivalent to the initial dynamics. The relation between $\Lambda$ and $(p,q)^T$ is embedded in the $R$-dependent matrix $U(R)$.

In fact, the above transformation is in accord with the quantum-form representation transformation to energy-representation. Because $\hat{H}(R)$ is doublet degenerate which determines a 2-dimensional degeneracy patch in classical phase-space, there are two zeros in $U \Gamma U^{-1}$ which defines two orthogonal directions on this patch. When the deviation is on the patch, the temporal evolution of the wavefunction vanishes (omitting the overall phase), $\dot{p}_i=\dot{q}_i=0$. One can readily generalize that, for the $m$-folder degeneracy, the degeneracy regime is $2(m-1)$ dimensional.

Thus the dynamics for the deviation reads,
\begin{equation}
U\left(\begin{array}{c}\frac{d \delta p}{dt}\\ \\
\frac{d \delta q}{dt}\end{array}\right)=\Gamma_{\text{dia}}\delta\Lambda-U\left(\begin{array}{c}\frac{\partial \bar{p}(R)}{\partial R} \\ \\
\frac{\partial \bar{q}(R)}{\partial R} \end{array}\right)\dot{R}.
\end{equation}
Expressed all by new variables, the above equation becomes,
\begin{equation} \label{firstE}
\frac{d\delta\Lambda}{dt}=\Gamma_{\text{dia}}\delta\Lambda-\frac{\partial\bar{\Lambda}}{\partial R}\dot{R}
+\frac{dU}{dR}U^{-1}\delta\Lambda\dot{R},
\end{equation}
where
\begin{equation}
\frac{\partial\bar{\Lambda}}{\partial R}=U\left(\begin{array}{c}\frac{\partial \bar{p}(R)}{\partial R} \\ \\
\frac{\partial \bar{q}(R)}{\partial R} \end{array}\right)\neq\frac{\partial}{\partial R}\left[ U\left(\begin{array}{c} \bar{p}(R) \\ \\
\bar{q}(R)\end{array}\right)\right],
\end{equation}

In Eq.~(\ref{firstE}), the last term on the right hand side is at least second-order with respect to $\dot{R}$ which is negligible in the first-order theory; the second term on the right is tightly associated with the Wilczek-Zee phase since $\bar{p}$ ($\bar{q}$) is defined as the variable predicted by Wilczek-Zee theory. At this stage, one may naturally ask what this term would be like in the current framework of phase-space of classical-kind. To answer this question, employing the original Wilczek-Zee formula becomes compulsory. For doublet degeneracy, the Wilczek-Zee theory predicts the wavefunction during the adiabatic evolution as,
\begin{equation}
|\psi\rangle=c_1(R)|D_1(R)\rangle+c_2(R)|D_2(R)\rangle,
\end{equation}
with $c_1(R)$ and $c_2(R)$ satisfying,
\begin{equation} \label{WZ}
\frac{d}{dR}\left(\begin{array}{c} c_1\\ \\
 c_2\end{array}\right)=-\left(\begin{array}{cc}\langle D_1|\frac{\partial}{\partial R}|D_1\rangle &\langle D_1|\frac{\partial}{\partial R}|D_2\rangle
\\ \\ \langle D_2|\frac{\partial}{\partial R}|D_1\rangle&\langle D_2|\frac{\partial}{\partial R}|D_2\rangle
\end{array}\right)\left(\begin{array}{c} c_1\\ \\
 c_2\end{array}\right).
\end{equation}
Considering now an infinitesimal segment of the adiabatic process $dR$. After each $dR$, the difference of the final wavefunction and the initial one according to (\ref{WZ}) reads,
\begin{eqnarray} \nonumber
|\psi(R+dR)\rangle-|\psi(R)\rangle=c_1\frac{\partial|D_1\rangle}{\partial R}dR+c_2\frac{\partial|D_2\rangle}{\partial R}dR \\ \nonumber -c_1\langle D_1|\frac{\partial}{\partial R}|D_1\rangle dR|D_1\rangle-c_2\langle D_1|\frac{\partial}{\partial R}|D_2\rangle dR|D_1\rangle \\
-c_1\langle D_2|\frac{\partial}{\partial R}|D_1\rangle dR|D_2\rangle-c_2\langle D_2|\frac{\partial}{\partial R}|D_2\rangle dR|D_2\rangle,
\end{eqnarray}
which can be easily proven to be orthogonal with any differential of the wavefunction on the degeneracy patch, $d|\psi\rangle=dc_1|D_1\rangle+dc_2|D_2\rangle$,
\begin{equation} \label{WZO}
\left(\langle\psi(R+dR)|-\langle|\psi(R)|\right)\cdot d|\psi\rangle=0.
\end{equation}
Equation (\ref{WZO}) implies that, after an infinitesimal time interval, the change of Wilczek-Zee state is orthogonal with any infinitesimal state defined on the degeneracy patch. In the language of phase-space, the projection of $\frac{\partial\bar{\Lambda}}{\partial R}$ on the degeneracy patch vanishes. Employing the same gauge as in (\ref{firstdia}), this term assumes the form,
\begin{equation} \label{firstWZ}
\frac{\partial\bar{\Lambda}}{\partial R}=U\left(\begin{array}{c}\frac{\partial \bar{p}(R)}{\partial R} \\ \\
\frac{\partial \bar{q}(R)}{\partial R} \end{array}\right)=\left(\begin{array}{c}0 \\ 0 \\
A_1 \\A_2 \\\vdots \\A_{2(n-2)} \end{array}\right).
\end{equation}

Combining Eq.~(\ref{firstdia}) and (\ref{firstWZ}) and neglecting the third term on the right hand side of Eq.~(\ref{firstE}), one can finally derive the dynamics for the deviation within the first-order approximation.
\begin{equation}
\frac{d\delta P_1}{dt}=\frac{d\delta Q_1}{dt}=0,
\end{equation}
which reveals the fact that, during degenerate adiabatic evolution, the deviation in the first-order approximation is always perpendicular to the degeneracy patch, with the projection of first-order state on the degeneracy patch satisfying the Wilczek-Zee theory. This property can be qualitatively explained: to fulfill the adiabatic following which is dictated by adiabatic theorem, a general driving force impelling the system along the adiabatic path must be present. Suppose during the adiabatic process there is no deviation from fixed point (eigenstate), the dynamics for the quantum state, described by $p$ and $q$ defined in (\ref{definition}), will always be,
\begin{equation} \label{classical0}
\quad \frac{d p_i}{dt}=-\frac{\partial H(R)}{\partial q_i}\equiv0\,,
~~~\frac{d q_i}{dt}=\frac{\partial H(R)}{\partial p_i}\equiv0,
\end{equation}
giving rise to a constant $p$ and $q$ (quantum state). On the other hand, according to the adiabatic theorem, the state should not be a constant but follow the Wilczek-Zee path during the adiabatic evolution. This paradox legalizes the emergence of the intrinsic adiabatic deviation. However, in the case of degeneracy levels, the deviation in the degeneracy subspace cannot induce a force (all the states on the subspace are dynamical fixed point), i.e., Eq.~(\ref{classical0}) still holds. Thus it is natural that the deviation prefers to be perpendicular to the degeneracy patch.

This result is in sharp contrast to the previous result of first order deviation~\cite{degenerate1,degenerate3} obtained by quantum adiabatic perturbation theory~\cite{degenerate}, where the projection of first-order deviation onto the degeneracy subspace is not zero. This contradiction might arise from the fact that the ansatz of the evolving wavefunction taken in~\cite{degenerate1} has reduced the Hilbert space and is thereby insufficient to describe all the possible states during degenerate adiabatic evolution, i.e., the first-order state derived here falls out of the ansatz taken in~\cite{degenerate1}.

The above treatment can be naturally generalized to the cases of higher degeneracy degrees. The only difference is that there are $2\times(k-1)$ zero diagonal elements of $\Gamma_{\text{dia}}$ shown in (\ref{firstdia}) and the same number of zero-element of vector shown in (\ref{firstWZ}), with $k$ being the degree of degeneracy.

In fact, as in the case of non-degenerate case, the deviation vertical to the patch will generally oscillate. To evaluate the deviation, it is convenient to take the nonzero part of $\Gamma_{\text{dia}}$ and $\frac{\partial\bar{\Lambda}}{\partial R}$ associated with the dynamics perpendicular to the degeneracy subspace,
\begin{equation} \label{firstNZ1}
\Gamma_{\text{dia}}^{\text{NZ}}=\left(\begin{array}{cccc}
d_1 & 0 & \ldots & 0\\
0  &d_2& \ldots & 0 \\
\vdots  &\vdots& \ddots & \vdots \\
0  & 0 &  \ldots    &d_{2(n-2)}     \end{array}\right);
\end{equation}
\begin{equation} \label{firstNZ2}
\left(\frac{\partial\bar{\Lambda}}{\partial R}\right)^{\text{NZ}}=\left(\begin{array}{c}
A_1 \\A_2 \\\vdots \\A_{2(n-2)} \end{array}\right).
\end{equation}
The dynamics for the deviation then reads,
\begin{equation} \label{firstNZD}
\frac{d\delta\Lambda^{\text{NZ}}}{dt}=\Gamma_{\text{dia}}^{\text{NZ}}\left[ \delta\Lambda-\Gamma_{\text{dia}}^{\text{NZ},-1}\left(\frac{\partial\bar{\Lambda}}{\partial R}\right)^{\text{NZ}}\dot{R}
\right],
\end{equation}
which clearly shows that the deviation orthogonal to the degeneracy region behaves like a multi-dimensional harmonic oscillator. The averaged deviation, proportional to the adiabatic speed $\dot{R}$, reads,
\begin{equation} \label{fdeviation}
\delta\Lambda^{\text{NZ}}=\Gamma_{\text{dia}}^{\text{NZ},-1}\left(\frac{\partial\bar{\Lambda}}{\partial R}\right)^{\text{NZ}}\dot{R},
\end{equation}
and the first-order deviation in the representation of $\Lambda$, according to our convention (\ref{conventionn}), should be written as,
\begin{equation}
\delta^1\Lambda=\left(\begin{array}{ccc}
0 & 0 & (\delta\Lambda^{\text{NZ}})^T \end{array}\right)^T,
\end{equation}
which shows clearly again that the first-order deviation vanishes in the degeneracy subspace.

As mentioned above, the relation between vector $\Lambda$ and the initial variables $(p_i,q_i)$ is embedded in $U$, i.e.,
\begin{equation}
\left(\begin{array}{c} \delta^1 p\\ \\
 \delta^1 q\end{array}\right)=U^{-1}\delta^1\Lambda,
\end{equation}
from which the first-order wavefunction can be obtained from the definition in Eq.~(\ref{definition}).

Because the first-order adiabatic deviation behaves like a harmonic oscillator, it forms another Hamiltonian dynamics as that in the non-degenerate case~\cite{Zhang}. As the center of the oscillator depends on $\dot{R}$, the first-order deviation will undergo a tiny adiabatic evolution as both $R$ and $\dot{R}$ evolves slowly, which is identical with the situation in non-degenerate adiabatic process.

\section{Dynamics for the degenerate adiabatic deviation: high-order theory}

In the last section, the first-order deviation is shown to be orthogonal to the degeneracy subspace. We in this section give the general formula for the high-order deviation.

Returning back to Eq.~(\ref{firstE}), the deviation $\delta\Lambda$ is in fact the sum of all orders of the deviations,
\begin{equation}
\delta\Lambda=\delta^1\Lambda+\delta^2\Lambda+\ldots .
\end{equation}
The dynamics of $\delta\Lambda\equiv\delta P_i,\delta Q_i$ is in fact the Tailor expansion instead of the first-order approximation associated with $\Gamma_{\text{dia}}$. However, the higher-order terms in the expansion is only associated with $\Gamma_{\text{dia}}^{\text{NZ}}$ since the deviation on the degeneracy patch can never generate any driving force.

First, let's consider the second-order term in Eq.~(\ref{firstE}). The dynamics for the second-order deviation then reads,
\begin{equation} \label{o2}
\frac{d\delta^2\Lambda}{dt}=\frac{1}{2}\delta\Gamma_{\text{dia}}\delta^1\Lambda
+\frac{dU}{dR}U^{-1}\delta^1\Lambda\dot{R},
\end{equation}
where $\delta\Gamma_{\text{dia}}$ is defined as
\begin{eqnarray} \label{shiftgamma}  \nonumber
\delta\Gamma_{\text{dia}}&=&\sum_i\left(\frac{\partial\Gamma_{\text{dia}}}{\partial P_i}\right)_{\bar{p},\bar{q}}\delta P_i+\sum_i\left(\frac{\partial\Gamma_{\text{dia}}}{\partial Q_i}\right)_{\bar{p},\bar{q}}\delta Q_i \\
&\equiv &\left(\left(\frac{\partial\Gamma_{\text{dia}}}{\partial P}\right)_{\bar{p},\bar{q}}, \left(\frac{\partial\Gamma_{\text{dia}}}{\partial Q}\right)_{\bar{p},\bar{q}}\right)\cdot\delta^1\Lambda
\end{eqnarray}
which has the same matrix dimension as $\Gamma_{\text{dia}}$. 

Next, according to Eq.~(\ref{firstE}), the dynamics for the $k$th order deviation can be iteratively obtained as,
\begin{equation} \label{ok}
\frac{d\delta^k\Lambda}{dt}=\sum_{j=1}^{k-1}\left(\Delta^j\Gamma_{\text{dia}}\right)\delta^{k-j}\Lambda
+\frac{dU}{dR}U^{-1}\delta^{k-1}\Lambda\dot{R},
\end{equation}
with
\begin{equation} \label{long}
\Delta^j\Gamma_{\text{dia}}={\cal T}^j\left\{\sum_{i=1}^j \frac{1}{(i+1)!} \left[\left(\frac{\partial}{\partial P},\frac{\partial}{\partial Q}\right)\cdot\sum_{r=1}^j\delta^r\Lambda \right]^i\Gamma \right\}
\end{equation}
The function ${\cal T}^j(\ldots)$ in (\ref{long}) is to take the $j$th-order terms in $(\ldots)$

Because $dU/dR$ is generally very different from $U$ itself, the deviation with the order higher than one will not be zero on the degeneracy subspace. However, if the normal direction of the degeneracy subspace in the whole Hilbert-phase space keeps fixed as parameter $R$ changes, $dU/dR$ vanishes and the last term in Eq.~(\ref{firstE}) is always zero, which means that in this case deviations of all orders are vertical to the degeneracy subspace and the formulation reduce to that for the non-degenerate case~\cite{Zhang}. This sheds more light on the difference between the degenerate and non-degenerate adiabatic evolutions.

As seen iteratively from Eqs.~(\ref{o2}) and (\ref{ok}), arbitrary order deviation evolves dynamically like a harmonic oscillator, with the center of the $k$th order depending on the temporal derivatives of $R$ up to the $k$th order. This situation is identical with that in the non-degenerate case.

\section{Numerical simulations}

\begin{figure}[t]
\begin{center}
\vspace*{-0.8cm}
\par
\resizebox *{8cm}{6cm}{\includegraphics*{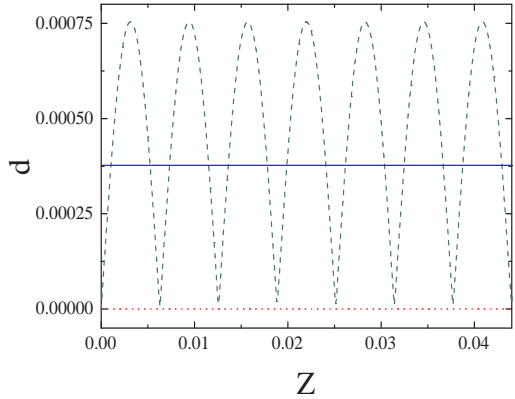}}
\end{center}
\par
\vspace*{-0.5cm} \caption{(color online) Numerical adiabatic deviations from the Wilczek-Zee theory for the tripod scheme Hamiltonian as $z$ is scanned with velocity $v=0.001$ and $x=1$. The dashed green curve and dotted red one are the results of deviations in the perpendicular direction and in the degeneracy subspace, respectively, when the initial state is set on $|D_2\rangle$; with the initial state set according to Eq.~(\ref{fdeviation}), the blue solid line is for the perpendicular deviation. Throughout $x$ and $z$ are in units of $1/k_l$, $v$ in units of $\Omega_0/\hbar k_l$. }\label{fig4}
\end{figure}

To verify our results, we employ the tripod scheme Hamiltonian implemented by three laser beams interacting with Rubidium atom. For convenience we adopt the same configuration as in Ref.
\cite{Juzeliunas2008PRL}, where two laser beams are counter-propagating along the $x$-axis and the third laser beam is
along the $z$-axis. The associated Hamiltonian under RWA is given by
\begin{equation}
H_{4}=\sum_{n=1}^{3}\Omega_{n}|0\rangle\langle n| + h.c.,
\end{equation}
with
\begin{eqnarray}
\Omega_{1}&=&\frac{\Omega_{0}\sin(\xi)}{\sqrt{2}}e^{-ik_lx}, \\
\Omega_{2}&=&\frac{\Omega_{0}\sin(\xi)}{\sqrt{2}}e^{ik_lx},\\
\Omega_{3}&=&\Omega_{0}\cos(\xi)e^{ik_lz},
\end{eqnarray}
where the parameter $\xi$ is set to satisfy $\cos(\xi)=\sqrt{2}-1$,
as in Ref. \cite{Juzeliunas2008PRL}, and $k_l$ is the wavevector of the laser fields.

The Hamiltonian $H_{4}$ has two degenerate states with a
null eigenvalue.  We denote these two degenerate states as
$|D_{1(2)}\rangle$ and it is straightforward to find their spatial
dependence as follows \cite{Juzeliunas2008PRL},
\begin{eqnarray}
|D_1\rangle&=&(|\tilde{1}\rangle-|\tilde{2}\rangle)e^{-i\kappa'
z}/\sqrt{2} \nonumber \\
|D_2\rangle&=&\left[\cos(\xi)\left(|\tilde{1}\rangle+|\tilde{2}\rangle\right)/\sqrt{2}
-\sin(\xi)|3\rangle\right]e^{-i\kappa' z},
\end{eqnarray}
where \begin{eqnarray} \kappa'&\equiv& k_l[1-\cos(\xi)],\\
|\tilde{1}\rangle &\equiv & |1\rangle e^{ik_l(x+z)}, \\
|\tilde{2}\rangle &\equiv& |2\rangle e^{-ik_l(x-z)}.
\end{eqnarray}

In the numerical simulation, we consider two scenarios: the quantum state emanates from (i) degeneracy subspace and (ii) the state predicted by Eq.~(\ref{fdeviation}). Then the parameter $x$ or $z$ is scanned as in the tripod scheme and calculate (1) the distance between real state derived by numerically integrating the Schr\"odinger equation and its projection state in degeneracy subspace and (2) the distance between projection state and the state obtained by Wilczek-Zee theory. According to our results, the former, which stands for the deviation in vertical direction, should be a quantity of first-order of $\dot{z}(\dot{x})$ while the latter, which stands for the deviation in the degeneracy subspace,  should be at least second-order. The typical results displayed in Fig.~2 as well as other numerical results clearly demonstrate this property, which verify our theory numerically. Here the distance between two states $|\psi_1\rangle$ and $|\psi_2\rangle$ is defines as $d=\sqrt{(\langle\psi_1|-\langle\psi_2|)\cdot(|\psi_1\rangle-|\psi_2\rangle)}$.

\section{Conclusion}

In summary, the deviation during quantum adiabatic evolution for degeneracy energy levels is studied both analytically and numerically. In the first-order formulation with respect to adiabatic speed, the deviation between exact state and Wilczek-Zee state will always be in the direction perpendicular to the degeneracy subspace. Thus the deviation in the degeneracy subspace will be at least second-order. Our findings are of fundamental interest to non-Abelian quantum computation and topological braiding. The implications of this work for designing optimal protocols of degenerate adiabatic quantum gate should be a fascinating topic in our future studies.

ACKNOWLEDGMENTS

Thanks to Biao Wu from Peking University for helpful discussion.

\bibliographystyle{apsrev}

\end{document}